\documentclass[pra,preprint]{revtex4-1}
\usepackage{CJK}
\usepackage{graphicx}
\usepackage{amsmath}
\usepackage{amssymb}
\usepackage{amsmath}
\usepackage{color}
\usepackage{bm}
\usepackage{latexsym}
\usepackage{hyperref}
\usepackage{graphicx}
\usepackage{epstopdf}
\usepackage{subfigure}
\usepackage[normalem]{ulem}
\usepackage{tikz}
\usetikzlibrary{positioning}
\newcommand{\void}[1]{}

\newcommand{\da}{\dagger}
\newcommand{\non}{\nonumber}

\newcommand{\be}{\begin{equation}}
\newcommand{\ee}{\end{equation}}

\begin{document}
\title{Revealing quantum effects in bosonic
Josephson junctions: a multi-configuration
atomic coherent states approach}

\author{Yulong Qiao}
\affiliation{Institut f\"ur Theoretische Physik, Technische Universit\"at Dresden, 01062 Dresden, Germany}

\author{Frank Grossmann}
\email{frank.grossmann1@tu-dresden.de}
\affiliation{Institut f\"ur Theoretische Physik, Technische Universit\"at Dresden, 01062 Dresden, Germany}



\date{\today}
\begin{abstract}
The mean-field approach to two-site Bose-Hubbard systems is well
established and leads to nonlinear classical equations of motion
for the population imbalance and the phase difference. It can,
e.\ g., be based on the representation of the solution of the time-
dependent Schr\"odinger equation either by a single Glauber state
or by a single atomic (SU(2))
coherent state [S. Wimberger et al., Phys. Rev. A {\bf
103}, 023326 (2021)]. We demonstrate that quantum effects beyond
the mean-field approximation are easily uncovered if, instead,
a multi-configuration
ansatz with a few time-dependent SU(2) basis functions is used in
the variational principle. For the case of plasma oscillations, the
use of just two basis states, whose time-dependent parameters are
determined variationally, already gives good qualitative agreement
of the phase space dynamics with numerically exact quantum
solutions. In order to correctly account for more non-trivial
effects, like macroscopic quantum self trapping, moderately more
basis states are needed. If one is interested in the onset of
spontaneous symmetry breaking, however, a multiplicity of two gives
a big improvement towards the exact result already. In any case, the number of
variational trajectories needed for good
agreement with full quantum results is orders of magnitude smaller
than in the semiclassical case, which is based on multiple mean-field
trajectories.

\end{abstract}
\maketitle

\section{Introduction}

The Bose-Hubbard (BH) model of $S$ interacting (bosonic) atoms in optical lattices is
the basis of many state of the art experimental \cite{Getal02,BDZ08,PSSV11,Tetal12} as
well as theoretical efforts \cite{JZ05,PSSV11,Kosl16}. The cold atom Hubbard tool box
introduced in \cite{JZ05} puts a focus on strongly interacting
many-body dynamics and embraces the fields of quantum optics, quantum computation and
solid state physics. The BH model is a paradigm for the
rich physical phenomena exhibited in these areas, such as there are
quantum phase transitions between the superfluid and the Mott insulator phase \cite{Getal02}, self-trapping in bosonic Josephson junctions
\cite{MCWW97}, and quantum chaology \cite{Kosl16}, to name just a few.

Restricting the amount of lattice sites makes the
quantum dynamics of the BH model easily tractable numerically
for moderate particle numbers. Recent theoretical work has thus focused on the
cases of four (and six) sites \cite{TPUUR18} with different levels of
approximation: exact, semiclassical and classical (mean-field, or
truncated Wigner approximation (TWA)).
Also the trimer (ring) case has been studied, due to the facts that it
is leading to the melting of discrete vortices via quantum
fluctuations \cite{LAK06} and that it is the smallest
system that displays a mixed phase-space mean-field dynamics
without an external driving term \cite{AVC14,NaHa23}.
This system has also been dealt with using a
group theoretical \cite{NHMM00,FrPe01} and a semiclassical time-domain approach \cite{SiSt14}. With an additional drive (periodic kicks) even
the double well case is showing signatures of chaos \cite{KCV13}.
Furthermore, the case of two wells without external driving has been
extensively studied. The system dynamics has, e.\ g., been investigated both
in a mean-field classical approximation and  fully (and perturbatively) quantum mechanically \cite{MCWW97,TLF05,STFL06,Java10,FTS22,SULT22},
as well as also semiclassically, using a phase space picture \cite{Chuetal10}, or employing the Herman-Kluk propagator
\cite{HK84,SiSt14}. This same propagator has also
been used in a semiclassical time-domain study of the single well problem \cite{jpa16}. Furthermore, the driven single well problem has served
as a model in a study of dynamical tunneling \cite{WDWD12}.

An important lesson from the vast literature is that semiclassical
approaches do well in reproducing the full quantum results,
while the mean-field and/or truncated Wigner approach have their
limitations. TWA does, e.g.,
not allow for the investigation of revival phenomena, present in the
quantum dynamics \cite{SULT22}. In contrast, the macroscopic quantum
self trapping effect in bosonic Josephson junctions could already be uncovered using a mean-field approach based on the Gross-Pitaevskii equation \cite{MCWW97}.
It turns out that mean-field theory predicts the transition to
macroscopic quantum self trapping at too large values of the on-site
interaction strength, however \cite{Wim21}.

In the following, we will focus on the quantum dynamics in the
case of two wells, for which the direct experimental observation
of tunneling and self trapping has become possible \cite{Albiez05}.
Theoretically, this case has been reviewed in
\cite{Legg01} as well as in \cite{BaFo16}, where the exact solubility
of the eigenvalue problem in terms of the Bethe-Ansatz has been reviewed.
Furthermore, a fresh look on finite size (i.e., finite particle number)
effects in the mean-field dynamics of those Josephson junction systems has
been given by Wimberger et al.\ \cite{Wim21}. These authors have used a
so-called atomic or SU(2) generalized coherent state \cite{ACGT72},
to uncover mean-field $1/S$ corrections to the more familiar mean-field
results based on standard Glauber coherent states. We will also employ those
favorable number conserving SU(2) states here. We will not use them in
a mean-field spirit, however, where just a single state is taken to solve
the time-dependent Schr\"odinger equation (TDSE).
In contrast, we will investigate what happens if we allow for non-trivial
multiplicity, which for reasons of simplicity we first choose to be just
two, i.\ e., we will use a superposition of two SU(2) states to solve
the TDSE. Inspired by previous experience with Gaussian-based
approaches to solve the TDSE for molecular Hamiltonians
\cite{irpc21,Zhao23}, as well as for spin-boson-type problems
\cite{jcp19-02,pra20,jcp22}, and due to the entanglement entropy studies
in \cite{LRP18} using two SU(2) states, we are  confident that only a
handful of suitable time-dependent basis states could
be enough to achieve satisfactory agreement with exact quantum solutions
if a full fledged variational approach is taken. In order to correctly
account for more demanding quantum effects like self trapping, it
will turn out that the multiplicity has to be increased, but it can still
be kept below the total number of time-independent Fock states that
has to be used in a full quantum calculation. Furthermore,
it is expected that the number of quantum
trajectories needed for convergence will be much reduced as
compared to semiclassical trajectory calculations that are based
on multiple mean-field trajectories.

The presentation is structured as follows: In order to set
the stage, in Sec.\ 2 we briefly review the mean-field approach,
based on a single atomic coherent state (ACS), to the dynamics of
the bosonic Josephson junction. At the end of this section, a
special focus will be put on the stability analysis of the nonlinear
classical phase space dynamics. In Sec.\ 3, we then choose an
ansatz wave-function with non-trivial multiplicity, employing
a small number of time-evolving atomic coherent states to
represent the quantum beat dynamics
(collapse and revival of the population imbalance)
of the BH dimer. In a brief review of the quantum phase operator concept, we
establish the relation between the phase difference in mean
field and its quantum analog. This allows us to compare numerical
results for phase space trajectories with the corresponding
solution of the TDSE. We will cover a broad range of system
parameters as well as initial conditions. It will turn out that
there are cases, close to the equilibrium point of the classical
dynamics, in which just two ACS will suffice to achieve reasonable
agreement with exact results. Away from the classical equilibrium,
the number of ACS will have to be increased, however.
In the last section we give conclusions and an outlook on possible
future work. Methodological details can be found in Section 5.

\section{Two-site BH model and mean field dynamics}

\subsection{The Hamiltonian}

The simplest Hamiltonian for the bosonic Josephson junction
(two-site BH model) in normal ordered form reads
\begin{equation}\label{eq:Ham}
\hat{H}=-J(\hat{a}_1^{\dagger}\hat{a}_2+\hat{a}_2^{\da} \hat{a}_1)
+\frac{U}{2}\sum_{j=1}^2\hat a_j^{\da 2}\hat a_j^2,
\end{equation}
where the bosonic ladder operators $\hat a_j$ and $\hat a^\dagger_j$ with
commutation relation $[\hat a_j,\hat a_j^\dagger]=\hat 1$ destroy, respectively create a particle (a bosonic atom) in the site labelled by the index $j$.
Furthermore,
\begin{equation}
\label{eq:pop}
 \hat n_j=\hat a_j^\dagger\hat a_j,\quad j=1,2
\end{equation}
counts the number of particles per site
and $\hat S=\hat n_1+\hat n_2$ is the total number operator and
its expectation value $S$ is a conserved quantity, because
$\hat S$ commutes with $\hat H$.

The (dimensionless) parameters $U$ and $J>0$ denote the strength
of the on-site interaction, determined by the s-wave scattering
length of the atomic species considered, and the tunneling amplitude,
respectively. Later on, we will consider positive and negative
values of $U$, corresponding to repulsive and attractive interaction
between the atoms, respectively.

\subsection{Mean-field dynamics}

The evolution of the BH model is governed by the
time-dependent Schr\"odinger equation (TDSE)
\begin{equation}
{\rm i}|\dot{\Psi}(t)\rangle=\hat{H}|{\Psi}(t)\rangle
\end{equation}
for the wave-function $|\Psi(t)\rangle$. Here as well as in the remainder
of this paper we have set $\hbar=1$.
In order to solve for the dynamics, in the present section
we will be following closely the mean-field work presented in \cite{Wim21}.

We start the discussion, by recalling the
eigenvalue equation of the annihilation operator in the form
\begin{equation}
\hat a_j|\alpha_j(t)\rangle=\alpha_j(t)|\alpha_j(t)\rangle
=\sqrt{n_j(t)}{\rm e}^{{\rm i}\phi_j(t)}|\alpha_j(t)\rangle,
\end{equation}
with the time-dependent Glauber coherent state \cite{Glauber} (displacement
operator applied to the ground state)
\begin{equation}
 |\alpha_j(t)\rangle={\rm e}^{-|\alpha_j(t)|^2/2}
 {\rm e}^{\alpha_j(t)\hat a^\dagger}|0\rangle
\end{equation}
and time-dependent average particle number $n_j(t)$ and
phase $\phi_j(t)$ of the site indexed by $j$. The position
space representation of this state is a
displaced Gaussian wavefunction \cite{irpc21}.

The approximate mean-field dynamics can then be obtained
by using an ansatz 
in terms of SU(2) coherent states (also refered to as atomic coherent states (ACS)
\cite{ACGT72}), defined by
\begin{eqnarray}
 |\Psi(t)\rangle\non&=&\frac{1}{\sqrt{S!}}
 \left(\sqrt{\frac{1+z(t)}{2}}\hat a_1^\dagger
 +\sqrt{\frac{1-z(t)}{2}}e^{-{\rm i}\phi(t)}\hat a_2^\dagger\right)^S
 |0,0\rangle\\
 &=&|S,\sqrt{\frac{1+z(t)}{2}},\sqrt{\frac{1-z(t)}{2}}e^{-{\rm i}\phi(t)}\rangle.
 \label{su2}
\end{eqnarray}
Here $|0,0\rangle$ is a shorthand notation for the direct product of
two single-particle vacuum states and the time-dependent parameters
\begin{equation}
z(t)=\frac{n_1(t)-n_2(t)}{S}
\end{equation}
and
\begin{equation}
\phi(t)=\phi_1(t)-\phi_2(t)
\end{equation}
are the (normalized) population imbalance and the relative phase of the two sites, respectively \cite{Wim21}. The
time-dependent particle number expectations at site $j$ can take on
fractional values.
As a simple example, the reader may want to consider
the case of $S=2$ and initial $z(0)=1/2$, for which $n_1(0)=3/2$
and $n_2(0)=1/2$.

If the system dynamics is governed by a harmonic oscillator
Hamiltonian or a Rabi model (single harmonic mode coupled to a
spin system), the use of the Glauber coherent
states mentioned above is common \cite{He913,pccp17}.
In the present case, we opt for using the generalized coherent
states (GCS) \cite{Pere,ZFG90}, which for two modes are
the SU(2) coherent states introduced above, instead of
a direct product of Glauber coherent states,
however. This is because the
former are better suited to describe particle
number conserving dynamics, as the latter consist of a
superposition of number states in the general case
\cite{TWK08,TWK09}. For Bose-Einstein condensates (BEC) this
observation has also been made by Schachenmayer et al.\
\cite{SDZ11}, who showed that the
multi-well Glauber coherent state ansatz is equivalent to the
GCS ansatz only in the case of large particle numbers. Furthermore, it is worthwhile to note that the highly entangled GCS is the ground state of the ``free-boson'' model, i.\ e., the BH model with vanishing on-site interaction, $U=0$ \cite{MCWW97,LDA12,MSPT11,LDA22}.

The representation of the GCS in the last line in Eq.\ (\ref{su2}) is motivated
by the general expression of a multi-mode generalized coherent state (total number of modes given by $M$) in the
form \cite{BP08}
\begin{equation}
\label{eq:GCS}
|S,\vec{\xi}\rangle=
\frac{1}{\sqrt{S!}}\Big(\sum_{i=1}^M\xi_ia_i^\dagger\Big)^S|0,0,\cdots,0\rangle,
\end{equation}
where the entries of the vector $\vec\xi$  are the complex parameters
$\{\xi_i\}$, which obey the ``normalization'' condition
$\sum_{i=1}^M|\xi_i|^2=1$. The representation of the unit operator in
these states has been used in \cite{pra21} to establish an exact variational
dynamics of the multi-mode Bose-Hubbard model. The number of independent real
parameters of the GCS in the two-site case, $M=2$, is three
(two complex numbers minus the normalization condition mentioned above)
but there is an overall phase factor that is irrelevant, however, so that
we just remain with the two real parameters $z$ and $\phi$ introduced above.
In the case of arbitrary site numbers, the equations for the  parameters $\xi_i$
are refered to as discrete nonlinear Schr\"odinger equation, which can be viewed as the discrete analog of the Gross-Pitaevskii equation for a BEC \cite{Kosl16}.

In the following, we will focus on the
Josephson junction case. The mean-field equations for the real parameters $z(t)$ and $\phi(t)$ are given by \cite{Wim21}
%
\begin{eqnarray}
 \dot{z}&=&2J\sqrt{1-z^2}\sin\phi:=f_1,
 \label{eq:mf1}
 \\
 \dot{\phi}&=&-2J\frac{z}{\sqrt{1-z^2}}\cos\phi-U(S-1)z:=f_2,
 \label{eq:mf2}
\end{eqnarray}
which are equations of motion of non-rigid pendulum type
\cite{SFGS97,PKSL01,GK07}. A stationary solution of these
coupled nonlinear equations is given by the equilibrium
points $(0,2\pi n)$ with $n=0,\pm 1,\pm 2,\dots$.

In the next step, we linearize the system of equations
around one of the equilibrium points. The Jacobian matrix
\cite{Wim2}  at $(z^\ast,\phi^\ast)=(0,0)$ is given by
\begin{equation}
\label{eq:Jac1}
 {\bf J}=\left(
 \begin{array}{cc}
  \left .\frac{\partial f_1}{\partial z}\right|_{z^\ast,\phi^\ast} &
  \left .\frac{\partial f_1}{\partial \phi}\right|_{z^\ast,\phi^\ast}
  \\
  \left .\frac{\partial f_2}{\partial z}\right|_{z^\ast,\phi^\ast} &
  \left .\frac{\partial f_2}{\partial \phi}\right|_{z^\ast,\phi^\ast}
 \end{array}
 \right)
 =\left(
 \begin{array}{cc}
  0 & 2J\\
  -2J-(S-1)U & 0
 \end{array}
 \right)
\end{equation}
and its eigenvalues are
\begin{equation}
\label{eq:ev1}
 \lambda_\pm=\pm\sqrt{2}J\sqrt{-2+\frac{U}{J}-\frac{US}{J}}.
\end{equation}
The so-called strength parameter
\begin{equation}
\label{eq:Lam}
\Lambda=U(S-1)/(2J)
\end{equation}
is an appropriate parameter combination to be used frequently in the
following. More details on the
linearized mean-field equations around the stationary points
can be found in Appendix \ref{app:plasma}.

\begin{figure}[h]
  \includegraphics[width=0.99\textwidth]{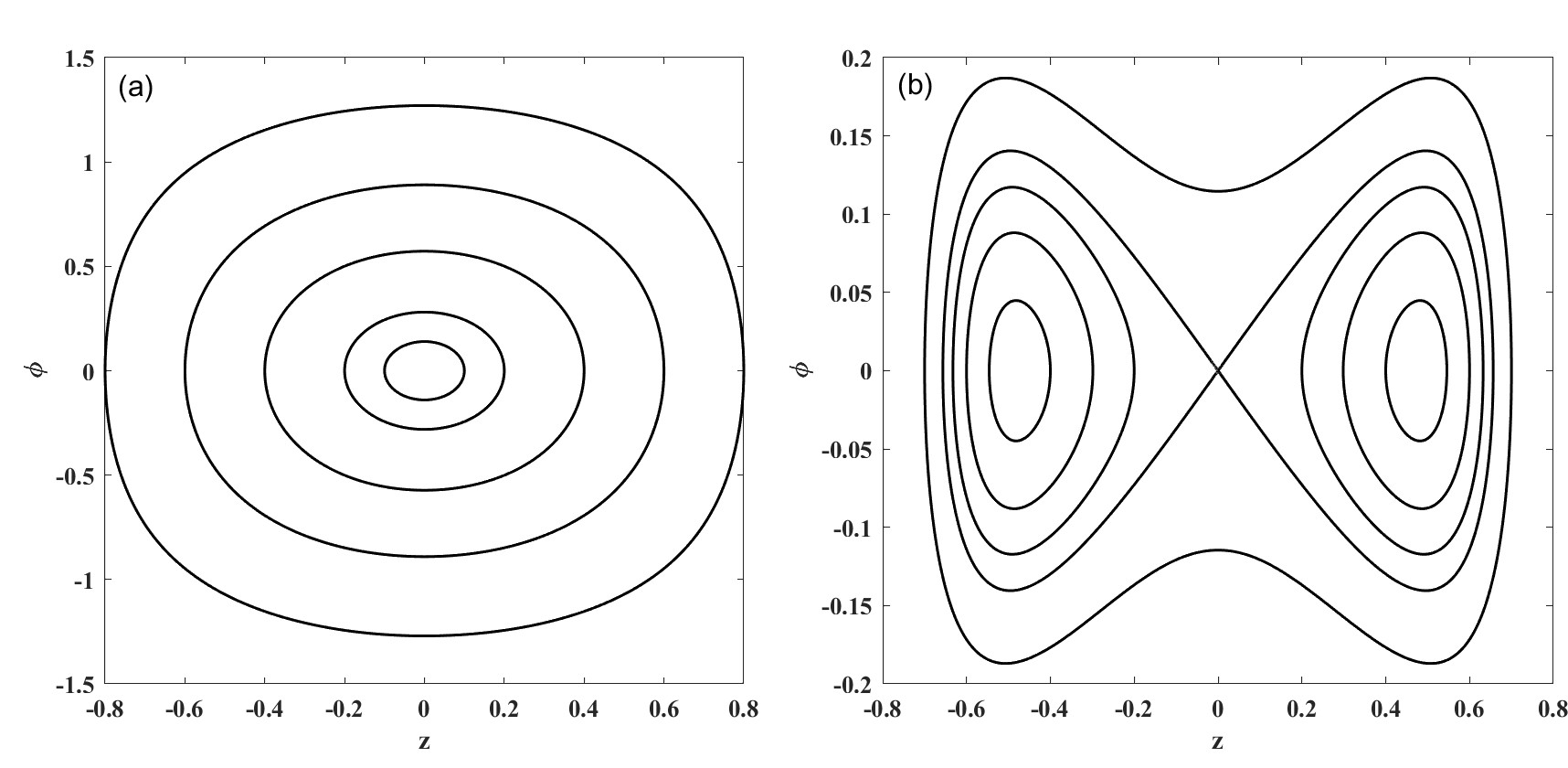}
  \caption{Phase space trajectories from the mean-field dynamics for different
  initial conditions and different values of on-site interaction strength:
  (a) $U/J=0.1$, (b) $U/J=-0.12$; the number of particles is $S=20$ in both cases.}\label{fig:mf}
\end{figure}

A qualitative change in the mean field dynamics will occur,
when the radicant in Eq.\ (\ref{eq:ev1}) changes sign, which
happens at the critical value $\Lambda_{\rm SSB}=-1$, where the
index SSB stands for spontaneous symmetry breaking \cite{MSPT11}.
If $\Lambda>-1$, both eigenvalues are imaginary, which indicates that the above
equilibrium point is a stable one and the solution is symmetric around
the origin, whereas $\Lambda<-1$ will lead to the emergence of another class of
stable equilibrium points. The symmetry breaking solutions are located around
the new stationary point(s)
\begin{equation}
\label{eq:SSB}
(z^{\rm SSB},\phi^{\rm SSB})=(\pm\sqrt{1-\frac{1}{\Lambda^2}},2\pi n)
\end{equation}
of the system of equations (\ref{eq:mf1},\ref{eq:mf2}),
where $n\in {\mathbb Z}$ \cite{Wim21}.
The corresponding Jacobi matrix is given by
\begin{equation}
\label{eq:Jac2}
 {\bf J}
 =\left(
 \begin{array}{cc}
  0 & \frac{2J}{\sqrt{\Lambda^2}}\\
  -2J\Lambda(1+\Lambda\sqrt{\Lambda^2}) & 0
 \end{array}
 \right)
\end{equation}
and its eigenvalues are
\begin{equation}
\label{eq:ev2}
 \lambda_\pm=\pm\frac{2J\sqrt{-\Lambda^4-\Lambda\sqrt{\Lambda^2}}}
 {\sqrt{\Lambda^2}}.
\end{equation}
If $\Lambda<-1$ these are imaginary and the solution of the
linearized equations around the SSB points are oscillatory.
Both cases are displayed in Fig.\ \ref{fig:mf}, with the left panel
showing motion around the stable fixed point for $U/J=0.1$
and the right panel showing the motions in case of $U/J=-0.12$,
where the stable fixed point
at the origin has turned into an instable one and new stable
fixed points appear at positive and negative values of $z$, see also \cite{ZEGO10}, for an experimental realization of this scenario. In Appendix \ref{app:plasma}, an analytic expression for the linearized solution in panel (b) of Fig.\ \ref{fig:mf}
at small values of $z$ and $\phi$  is given.

In \cite{Wim21} it is shown that the mean-field
prediction for the onset of SSB based on single Glauber coherent states fails dramatically at small particle numbers.
The mean-field result based on a single SU(2) coherent state does
better than the Glauber state prediction at small $S$ but
is not exact. Both mean field predictions reproduce the
full quantum result more faithfully at large values of $S$, however,
as can be seen in Fig.\ 4 of \cite{Wim21}.

A further conclusion that can be drawn from the mean-field equations
(\ref{eq:mf1},\ref{eq:mf2}) is the fact that the quantity
\begin{equation}
 E=\frac{US}{4}(S-1)z^2-JS\sqrt{1-z^2}\cos\phi
\end{equation}
is a constant of motion \cite{Wim21}. This leads to the existence of
a parameter regime, in which the imbalance cannot become zero during an oscillation cycle and therefore,
the average value of $z$ will be nonzero. The condition for this macroscopic
quantum self trapping (MQST) effect is $E(z(0),\phi(0))>E(0,\pi)=JS$. In
terms of the strength parameter introduced above, the onset of self
trapping is at \cite{RSFS99}
\begin{equation}
\label{eq:MQST}
 \Lambda_{\rm MQST}=\frac{1+\sqrt{1-z^2(0)}\cos\phi(0)}{z^2(0)/2},
\end{equation}
depending strongly on the initial position in phase space.
In contrast to the case of SSB, the mean field MQST effect
sets in at too large positive values of $U$ (repulsive interaction),
whereas mean field predicts the onset of SSB at too small values of $|U|$
\cite{Wim21}.

\section{Beyond mean field dynamics}

Due to the shortcomings of the mean-field approach for
$U\ne 0$, as there are the absence of collapses and revivals
of the population imbalance \cite{STFL06}, as well as failures
in the prediction of the onset of MQST as well as SSB \cite{Wim21}, we will
now go beyond mean field by employing a
multi-configuration ansatz for the solution of the TDSE.

We first give an explicit derivation of the equations
of motion followed by a brief review of the phase operator concept, which is needed to display our quantum results.
The parameter regimes of the results to be presented are put together in Table 1, from which it can be inferred that we
pick parameters bordering and inside the Josephson regime
($1<\Lambda< S^2$), intermediate between the Rabi and Fock regimes
\cite{Legg01}, for which $\Lambda\ll 1$ and $\Lambda\gg S^2$, respectively.
In the Josephson regime, the parameters we chose lead from
simple to more complex collapse and revival dynamics,
with increasing breathing amplitude, all the way to the phenomena of
MQST and SSB, mentioned in the previous section. The Rabi regime is considered to be the most trivial of the three commonly studied regimes, while the
Fock regime cannot be described reliably by our approach.
\begin{table}
\begin{center}
\begin{tabular}{c||c|c|c|c}
\\
section&3.3.1&3.3.2&3.4&3.5
\\
\hline
\\
$|U|/J$&$0.1$&$0.1$&$1.2$&$\geq 0.12$
\\
&&&$0.53$&
\\
\hline
\\
$(S,z(0))$&$(20,\ll 1)$&$(20,0.5)$&$(20,0.5)$&$(20,0.71)$
\\
&&$(50,0.5)$&$(50,0.5)$&$(50,0.83)$
\\
\hline
\\
phenomenon&PO&PO&MQST&SSB
\end{tabular}
\caption{Hamiltonian and initial state parameters to be investigated in detail
in Section 3. The initial phase was zero in all cases.
The acronyms stand for: PO: plasma oscillation, MQST: macroscopic quantum self
trapping, SSB: spontaneous symmetry breaking. }
\end{center}
\end{table}

\subsection{Equations of motion}

As a step towards exactness of the solution, we replace the wave-function of
Eq.~(\ref{su2}) by a linear combination of $N$ time-dependent SU(2)
coherent states, written as in the general SU($M$) case of Eq.\ (\ref{eq:GCS}),
leading to
\begin{equation}\label{multi_su2}
 |\Psi(t)\rangle=\sum_{k=1}^NA_k(t)|S,\xi_{k1}(t),\xi_{k2}(t)\rangle.
\end{equation}
We stress that all the parameters, compactly written as vectors
$\bm{A}$ (with $N$ entries) and $\bm{\xi}$ (with $2N$ entries),
are time-dependent and complex-valued.
Their (nonlinear) equations of motion, again derived from the TDVP,
in the general case of arbitrary multiplicity $N$ as well as site number $M$
have been given in matrix form in the appendix of \cite{pra21}.

For being self-contained, here, we explicitly review the
variational procedure for the two-well problem.
With the trial state from Eq.\ (\ref{multi_su2}) the
Lagrangian $L:={\rm i}\langle \Psi|\partial_t|\Psi\rangle-\langle\Psi|\hat H|\Psi\rangle$ takes the
explicit form
\begin{eqnarray}
L&=&\sum_{k,j=1}^NA_k^*\dot{A}_j\langle\vec{\xi}_k|\vec{\xi}_j\rangle+iS\sum_{k,j=1}^NA_k^*A_j\sum_{i=1}^2\xi_k^*\dot{\xi}_{ji}\langle\vec{\xi'_k}|\vec{\xi'_j}\rangle
\nonumber\\
&-&\sum_{k,j=1}^NA_k^*A_j\Big[-JS(\xi_{k1}^*\xi_{j2}+\xi_{k2}^*\xi_{j1})\langle\vec{\xi'_k}|\vec{\xi'_j}\rangle
\nonumber\\
&+&\frac{U}{2}S(S-1)\left(\xi_{k1}^{*2}\xi_{j1}^2+\xi_{k2}^{*2}\xi_{j2}^2\right)\langle\vec{\xi^{''}_k}|\vec{\xi^{''}_j}\rangle\Big].
\end{eqnarray}
The corresponding Euler-Lagrange equations are given by
\begin{equation}
\frac{\partial L}{\partial u_k^*}-\frac{\rm d}{{\rm d}t}\frac{\partial L}{\partial \dot{u}_k^*}=0,
\end{equation}
where $u_k$ denotes one element of the set
$\{A_k,\xi_{k1},\xi_{k2}\}$ of $3N$ complex valued parameters in
$|\Psi\rangle$. For the coefficients, this leads to the equations of motion
\begin{equation}\label{eq:Ak}
{\rm i}\sum_{j=1}^N\dot{A}_j\langle \vec{\xi_k}|\vec{\xi_{j}}\rangle+iS\sum_{j=1}^NA_j\sum_{i=1}^2\xi_{ki}^*\dot{\xi}_{ji}\langle\vec{\xi^{'}_k}|\vec{\xi^{'}_j}\rangle-\frac{\partial H}{\partial A_k^*}=0,
\end{equation}
where
\begin{equation}
 \frac{\partial H}{\partial A_k^*} =\sum_{j=1}^NA_j\Big[-JS(\xi_{k1}^*\xi_{j2}+\xi_{k2}^*\xi_{j1})\langle\vec{\xi^{'}_k}|\vec{\xi_j^{'}}\rangle+\frac{U}{2}S(S-1)\sum_{i=1}^2\xi_{ki}^{*2}\xi_{ji}^2\langle\vec{\xi_k^{''}}|\vec{\xi_j^{''}}\rangle\Big].
\end{equation}
For the coherent state parameters $\xi_{jm}$ $(m=1,2)$, we get
\begin{eqnarray}
\label{eq:par}
 &&{\rm i}S\Bigl[\sum_{j=1}^NA_k^*\dot{A}_j\xi_{jm}\langle\vec{\xi_k^{'}}|\vec{\xi_j^{'}}\rangle+\sum_{j=1}^NA_k^*A_j\dot{\xi}_{jm}\langle\vec{\xi_k^{'}}|\vec{\xi_j^{'}}\rangle
 \nonumber\\
 &+&(S-1)\sum_{j=1}^NA_k^*A_j\sum_{i=1}^2\xi_{ki}^*\dot{\xi}_{ji}\xi_{jm}\langle\vec{\xi_k^{''}}|\vec{\xi_j^{''}}\rangle\Bigr]-\frac{\partial H}{\partial \xi_{km}^*}=0,
\end{eqnarray}
where
\begin{eqnarray}
\frac{\partial H}{\partial \xi_{k1}^*}&=&\sum_{j=1}^NA_k^*A_j\Big[-JS\xi_{j2}\langle\vec{\xi_k^{'}}|\vec{\xi_j^{'}}\rangle-JS(S-1)(\xi_{k1}^*\xi_{j2}+\xi_{k2}^*\xi_{j1})\xi_{j1}\langle\vec{\xi_k^{''}}|\vec{\xi_j^{''}}\rangle
\nonumber\\
&+&US(S-1)\xi_{k1}^*\xi_{j1}^2\langle\vec{\xi_k^{''}}|\vec{\xi_j^{''}}\rangle+\frac{U}{2}S(S-1)(S-2)\sum_{i=1}^2\xi_{ki}^{*2}\xi_{ji}^2\xi_{j1}\langle\vec{\xi_k^{'''}}|\vec{\xi_j^{'''}}
\rangle\Big]
\end{eqnarray}
and an analogous equation for the second index being 2.

Some numerical tricks to solve the highly non-linear, implicit equations
of motion in Eqs.\ (\ref{eq:Ak},\ref{eq:par}) have been devised in \cite{prb20} for the case of Glauber coherent state basis functions.
Because of the restriction to $M=2$ of the site number in the present
investigation, at least for moderate particle numbers, the TDSE can
also be solved easily by an expansion of the wave-function in
(time-independent) Fock states, whose coefficients
fulfill a (numerically more well-behaved) system of coupled linear
first order differential equations.
The number of Fock states that are required is determined by the particle number via $S+1$. More details on the full quantum (Fock space) calculations,
whose results will be refered to as exact quantum results in the following, are given in Appendix \ref{app:exact}.

We stress that in all exact as well as beyond mean-field calculations to be presented, we take a single ACS as the
initial condition of the dynamics.
For the beyond mean-field calculations this means that a single element out of
the set $\{A_k\}$ is nonzero initially, whereas
all other elements will take nonzero values only in the course of time.

\subsection{Brief Review of Phase Operator Concept}

From the wave-function given by Eq.\ (\ref{multi_su2}), we can calculate the time-dependent site populations by taking expectation
values of the operators from (\ref{eq:pop}) and from this the imbalance $z$ between the two sites. For the analog of the relative phase $\phi$,
we use the quantum phase operator concept \cite{BP86,LAK06}, leading to
the expectation values
\begin{eqnarray}
 \langle\cos\hat\phi\rangle&=&\frac{\langle \hat a_2^\dagger \hat a_1
 +\hat a_2 \hat a_1^\dagger\rangle}
 {\sqrt{2\left<2\hat n_1\hat n_2+\hat n_1+\hat n_2\right>}}
 \label{eq:cosphi}
 \\
 \langle\sin\hat\phi\rangle&=&\frac{{\rm i}\langle a_1^\dagger a_2-a_2^\dagger a_1\rangle}
 {\sqrt{2\left<2\hat n_1\hat n_2+\hat n_1+\hat n_2\right>}}
 \label{eq:sinphi}
 \\
  \langle\sin^2\hat\phi\rangle&=&\frac{1}{2}-
  \frac{\langle (\hat a_2^\dagger \hat a_1)^2+(\hat a_2 \hat a_1^\dagger)^2\rangle}{{2\left<2\hat n_1\hat n_2+\hat n_1+\hat n_2\right>}}
  \label{eq:sinsq}
\end{eqnarray}
of the cosine and sine of the phase operator and its sine square.
In addition, the variance of the sine is defined by
\begin{equation}
\Delta(\sin\hat \phi):= \langle\sin^2\hat\phi\rangle-
\langle\sin\hat\phi\rangle^2.
\end{equation}
The normalization condition $\langle\sin^2\hat\phi+\cos^2\hat \phi\rangle=1$
and the expectation of $\cos^2\hat\phi$ from \cite{LAK06} have been used
to derive Eq.\ (\ref{eq:sinsq}).

The relation between the sine of the classical phase
variable, displayed in Fig.\ \ref{fig:mf} and the expectation of the
sine of the quantum phase is
\begin{equation}
 \langle\sin\hat\phi\rangle=\frac{S\sqrt{1-z^2}}{\sqrt{S(S-1)(1-z^2)+2S}}
 \sin\phi,
 \label{eq:rel}
\end{equation}
as can be derived by applying the operator in (\ref{eq:sinphi}) to an ACS.
For $S\to \infty$, the prefactor on the RHS of the above equation becomes
unity and the quantum and classical expressions become identical.
Furthermore, in \cite{LAK06} it has been shown that
the melting of coherence between the two sites is mirrored by the vanishing
of the expectation of $\cos\hat \phi$ and the occurrence of large
fluctuations of the corresponding variance.

In the following, we will focus on parameters on the border as well as
inside of the most interesting regime, the so-called Josephson regime, for
which the parameters fulfill the condition $1< US/(2J)\ll S^2$ \cite{Legg01}.
Depending on the initial conditions, beyond mean-field effects
can be observed in this case. In addition, we will also allow for negative
values of the strength parameter $\Lambda$ smaller than -1, in order to study
the SSB case and will employ large positive $\Lambda$ values close to the
(mean-field) MQST regime.

\subsection{Plasma oscillations}

In the following, we first consider the case of small on-site interaction.
In addition, also the initial imbalance shall first be small. In a second
step, this imbalance shall be large at $t=0$.

\subsubsection{Small initial population imbalance}

For small values of $U$, as well as of $z$, we only include two
ACS in the Ansatz in Eq.\ (\ref{multi_su2}), i.\ e., we use $N=2$.
Initially, $\bm{\xi}$ can be still be parameterized in analogy to
the procedure of the previous section by
$(\xi_{11},\xi_{12})=(\sqrt{\frac{1+z_1}{2}},\sqrt{\frac{1-z_1}{2}}
e^{-i\phi_1})$ and $(\xi_{21},\xi_{22})=(\sqrt{\frac{1+z_2}{2}},
\sqrt{\frac{1-z_2}{2}}e^{-i\phi_2})$. To highlight the changes that
the inclusion of
an additional basis state leads to, for the first SU(2) state, we use three
different initial conditions, namely
$z_{1,i}\in\{0.01,0.05,0.1\}, \phi_1=0, A_1=1$.
For the second SU(2) state, the initial values are identical and are
fixed as $z_2=0, \phi_2=2\pi/3, A_2=0$.
\begin{figure}[h]
\centering
\includegraphics[width=0.99\textwidth]{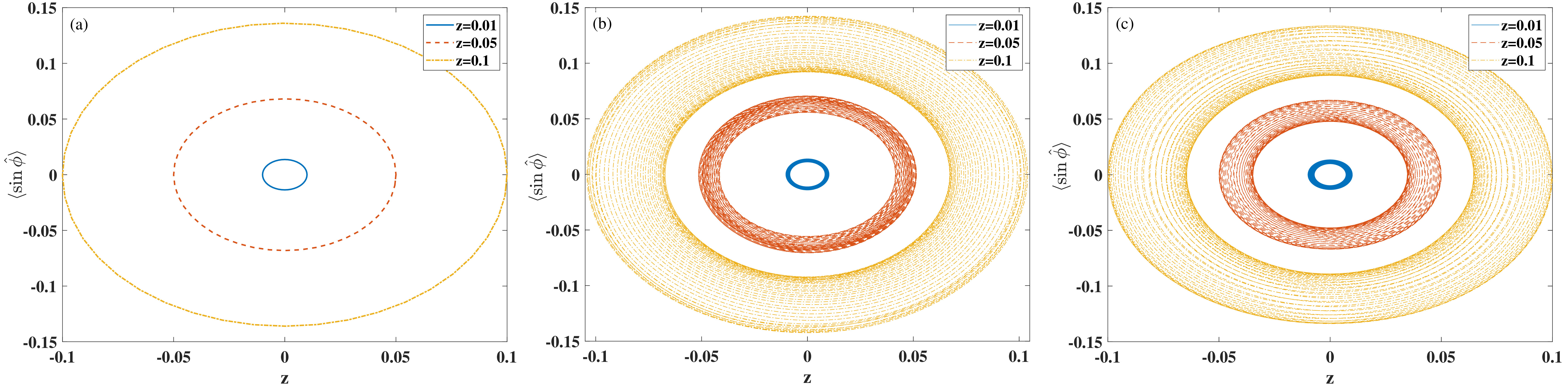}
\caption{Phase space trajectories in:  (a) mean-field, (b)
beyond mean field using ACS with $N=2$, and (c) exact quantum
dynamics. Different initial conditions are indicated by different
line styles: $z=0.01$ (solid blue), $z=0.05$ (dashed red), $z=0.1$
(dash-dotted yellow). System parameters are $U/J=0.1$ and $S=20$.}
\label{comp}
\end{figure}

The phase space trajectories in three different levels of
approximation and for the three different initial conditions
specified above are shown in Fig.\ \ref{comp}, for $S=20$ and
an on-site interaction strength of $U=0.1J$, implying $US/(2J)=1$.
We stress that the expectation value of the sine of the
phase operator is plotted on the $y$-axis. Its relation to the
sine of the phase difference in the mean-field case is given in
Eq.\ (\ref{eq:rel}).

In mean-field approximation all three initial conditions give
rise to an ellipsoidal phase space pattern as shown in the
previous section. Going beyond mean field by allowing for just one
additional SU(2), we see a qualitatively different behaviour which
corresponds to a  beating of the population imbalance, here
displayed by a spiraling
motion that first moves inward and then outward for all three
initial conditions. This is shown not to be an artefact by
comparison to the full quantum solution, which shows an almost
quantitative agreement with the ACS solution of multiplicity two.
We stress that the choice of the initial phase of the second
ACS is decisive for the quality of our beyond mean-field results.
Choosing $\phi_2$ to be zero, e.g., would lead to a spiraling in
the wrong direction.

In the case of the smallest imbalance displayed in Fig.\
\ref{comp}, the beating amplitude (the width of the blue ring)
is smallest and the description of the quantum dynamics with a
single classical (mean-field) trajectory
is almost adequate, as it would be in the Rabi-oscillation regime,
in which $\Lambda\ll 1$, a case, we are not considering herein.
As has been shown in \cite{SiSt14}, in order to cope with the
collapse and revival of the population imbalance oscillations, a
multitude of {\sl classical} trajectories is needed, however. A
ballpark number for the sample size in the Monte-Carlo integrations
performed by Simon and Strunz is 10$^4$. The truncated Wigner
approximation based on a similar sampling procedure does {\sl not}
capture the revival oscillations, but a full fledged semiclassical
approach is required to this end. To put our work in context, we
stress, that to capture the quantum behaviour displayed in Fig.\
\ref{comp} almost quantitatively, we need only two
``trajectories'', i.e., two ACS. This dramatic reduction
in basis size is due to the fact that in our present case, also the
trajectories (the dynamical evolution of the basis function
parameters) undergo the full variational procedure, i.e., they are
not mean-field trajectories. We are thus loosing the intuitive
appeal of a semiclassical method at the benefit of much smaller
computational effort, although the calculation of the quantum
trajectories is more involved than that of the mean-field ones.
In passing, we note that in a comparison of the hierarchy of
variational methods based on Glauber coherent states, applied to
the anharmonic Morse potential in \cite{irpc21}, the reduction in
basis function size was counteracted by the (numerical)
complexity of the solution of the variational equations of motion.

\subsubsection{Large initial population imbalance}

\begin{figure}[h]
\centering
\includegraphics[width=0.99\textwidth]{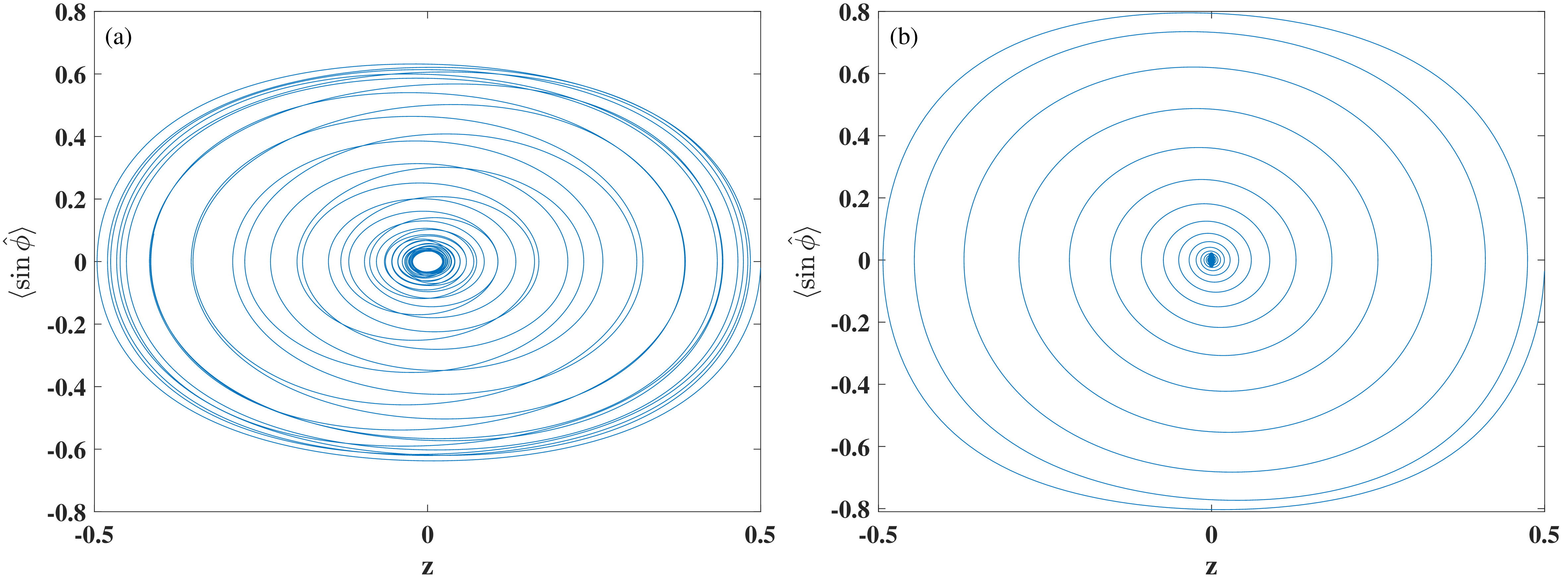}
\caption{Exact quantal phase space trajectories for times up to
$Jt=100$ in the case $U/J=0.1$ for (a) $S=20$ and (b) $S=50$. The
initial condition is $z=0.5$ in both cases.}\label{fig:z1_05}
\end{figure}

\begin{figure}[h]
\centering
\includegraphics[width=0.99\textwidth]{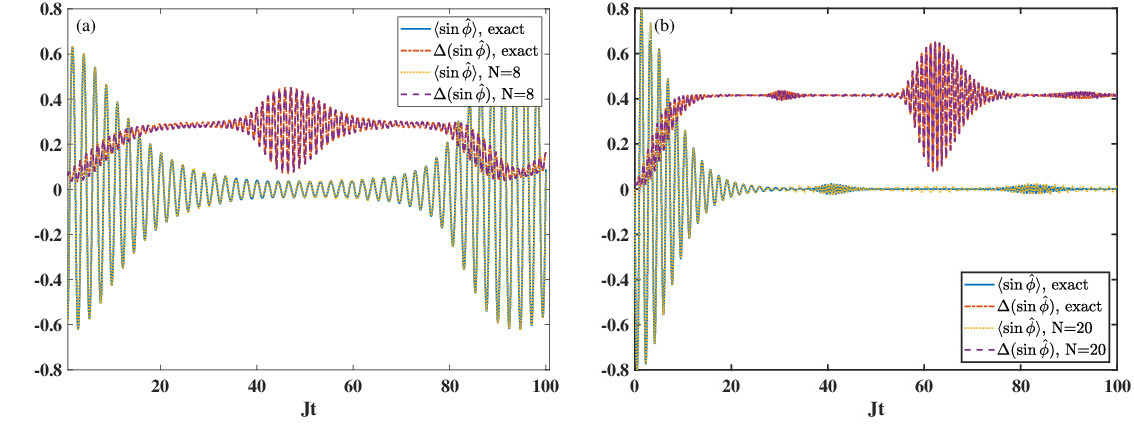}
\caption{A comparison of beyond mean-field and exact quantum
results for (a) $S=20$ and (b) $S=50$. The initial condition is
$z=0.5,\phi=0$ in both cases. The system parameter is $U/J=0.1$. The
expectation value of the sine of the phase (exact results: solid
blue line, multi ACS results: dotted yellow line) and its variance
(exact results: dash-dotted red line, multi ACS results: dashed
purple line) are displayed.}\label{fig:phase}
\end{figure}

The small initial imbalance in the previous case has led to an incomplete
collapse, i.e., the oscillation amplitude was still rather large in all three
cases at all times, with only small beating amplitude. In order to
suppress the total oscillation amplitude, i.e., to see very small
oscillations at least temporarily, we have to
allow for larger initial imbalances, which will be done next.
For the case of $U=0.1J$ and with initial $z=0.5$, the exact quantum dynamics
for two different total particle numbers, $S=20$ and $S=50$, is shown in
Fig.\ \ref{fig:z1_05}. Corresponding mean-field calculations (not shown)
would display a closed single loop oscillation without any spiraling in
(decrease of the oscillation amplitude). In the quantum case, however,
we see an almost complete collapse of the amplitude, the larger the
particle number. For the
larger $S$, in addition, the population imbalance as well as the
expectation of the sine of the phase operator
stay around zero for a longer time (see also  Fig.\ \ref{fig:phase}).

In Fig.\ \ref{fig:phase} we display the time evolution of the sine
of the phase operator and its variance as defined above.
The results of the beyond mean-field approach and exact
quantum calculations are compared. First, we observe the collapse
and revival in the case of $S=20$.
For $S=50$, the maximum time considered is too short to observe the
revival, however. Also we can see that the suppression of the oscillation amplitude of $\langle \sin\hat \phi\rangle$
comes along with an increase in the amplitude of variance oscillation.
Furthermore, agreement almost within line thickness between exact
and ACS results can be achieved, but only if the multiplicity is increased considerably compared to the previous case of small initial imbalance.
The multiplicities needed are $N=8$ in the case of $S=20$ and $N=20$
in the case of $S=50$.
The occurrence of large amplitude oscillations in the variance has to be
accounted for by an increase in the multiplicity, because in the single
ACS case the relative phase is well-defined.
We note that both multiplicities are smaller than the total number
of Fock states required, which is $S+1$. Furthermore, the choice of
the initial conditions for the initially unpopulated ACS is done
in the random fashion explained in detail in \cite{pra21}.

\subsection{Macroscopic quantum self trapping}

In order to observe self trapping in the Josephson regime, i.e., the restriction of the population dynamics, such that the  population on
one side is always larger than on the other side
the initial condition and/or the on-site interaction strength has to
be changed. From a mean-field argument the condition given in
Eq.\ \ref{eq:MQST} has been derived, which is valid at all times. In the following, we will use $z(0)=0.5$ and $\phi(0)=0$. This leads to
$\Lambda_{\rm MQST}\approx 15$.
We are choosing the total number of particles and the on-site interaction
strength such that the actual value of $\Lambda$ is just below the critical mean-field one, and that the classical dynamics therefore will not be trapped
but the strength parameter is large enough for the quantum trajectory
to be trapped at positive values of $z$ \cite{Wim21}.
For $S=20$, we take $U/J=1.2$ and for $S=50$, we take $U/J=0.53$,
leading to $\Lambda\approx 14.4$ and $\Lambda\approx 13.0$, respectively.
Both values are deep inside the Josephson regime.

\begin{figure}[h]
\centering
\includegraphics[width=0.99\textwidth]{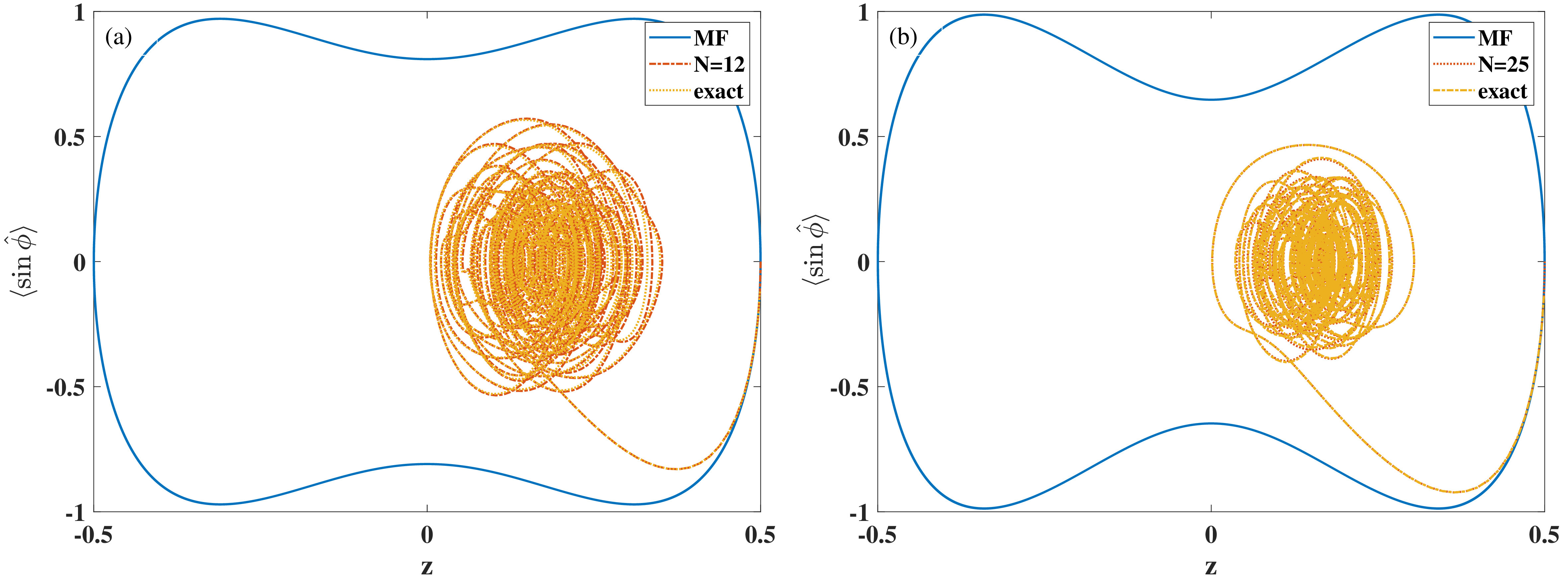}
\caption{Phase space trajectories for times up to $Jt=50$ in mean field approximation (solid blue) multi ACS (dash-dotted red) and exact quantum dynamics (dotted yellow). The initial condition is $(z(0),\phi(0))=(0.5,0)$. System parameters are: (a) $U/J=1.2$ and $S=20$ and (b) $U/J=0.53$ and $S=50$.}\label{fig:MQST}
\end{figure}

In Fig.\ \ref{fig:MQST}, the results for the phase space trajectories
followed up to a total time of $T=50J$ are displayed. As dictated by our choice
of parameters, the mean-field results do not display the MQST effect just yet.
The quantum MQST has set in already, however. The fact that in the exact
quantum results MQST happens for smaller coupling strengths than in mean field
has also been reported in \cite{Wim21}. There it was found that the
use of a single ACS does not allow one to observe this quantum effect
(the reduction of the critical $\Lambda$ value), however.
By consulting the red curves in
Fig.\ \ref{fig:MQST}, it can be seen that in order for the ACS-ansatz
to show the correct quantum behavior (early onset of MQST) a non-trivial
multiplicity has to be employed. In our present case this is $N=12$
for the case $S=20$ and $N=25$ for the case $S=50$. The high
multiplicities needed are due to the fact that both the initial condition
in $z$ and also $U$ are rather large. A single ACS will
only give the exact quantum solution for $U=0$, however.

\subsection{Spontaneous symmetry breaking}

So far, we have focused on positive on-site interaction strength.
It was already observed on the mean-field level, however,
that a spontaneous symmetry breaking is triggered by negative
values of $U$ beyond a certain threshold. The comparison of
the mean-field with the full quantum solution as well as our multi-configuration
ACS approach to this case will be in the focus of the present section.

\begin{figure}[h]
\includegraphics[width=0.99\textwidth]{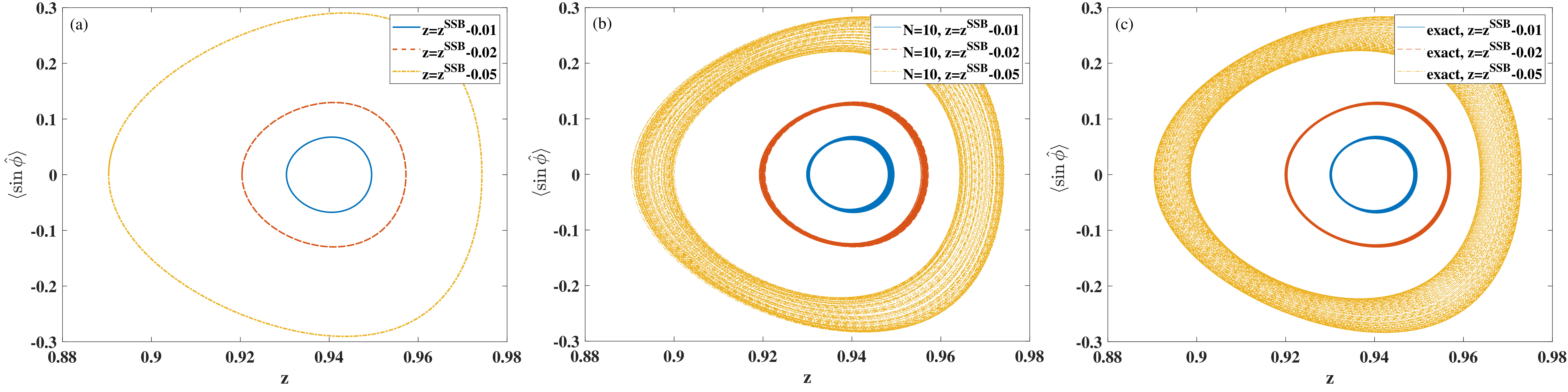}
\caption{Phase space trajectories for times up to $Jt=100$ in  (a)
mean-field, (b) beyond mean field with $N=10$ and (c) exact
quantum dynamics. Different initial conditions are displayed
by different line styles:
$z^{\rm SSB}-0.01$ (solid blue), $z^{\rm SSB}-0.05$ (dashed red),
$z^{\rm SSB}-0.1$ (dash-dotted yellow) with $z^{\rm SSB}=0.94$. Parameters are $U/J=-0.12$ and $S=50$.}\label{fig:plec}
\end{figure}

Because the mean-field prediction for SSB is good for large
particle numbers \cite{Wim21}, in Fig.\ \ref{fig:plec} we first consider the
case of $S=50$ and we take $U/J=-0.12$, leading to $\Lambda<-1$. Results of
three different levels of approximation are again displayed: mean-field, ACS
with small multiplicity (here $N=10$) and full quantum. In the mean-field case,
displayed in panel (a), we observe that the elliptic orbit for small deviations
from the symmetry breaking equilibrium point
$(z^{\rm SSB}\approx 0.94, \phi^{\rm SSB}=0)$, for larger displacements turns
into a  plectrum shaped orbit around the new stable fixed point (see also panel
(b) of Fig.\ \ref{fig:mf}).
As in the previous section, multi-configuration ACS with a small
multiplicity of $N=10$ displays the spiraling away from the mean-field
orbit (the ``quantum effect'') in very faithful way. The further away from
$z^{\rm SSB}$ the initial condition is, the broader the range of the
spiraling motion turns out to be, both in the ACS (panel (b)) and the exact
results (panel (c)).

\begin{figure}[h]
\centering
\includegraphics[width=0.99\textwidth]{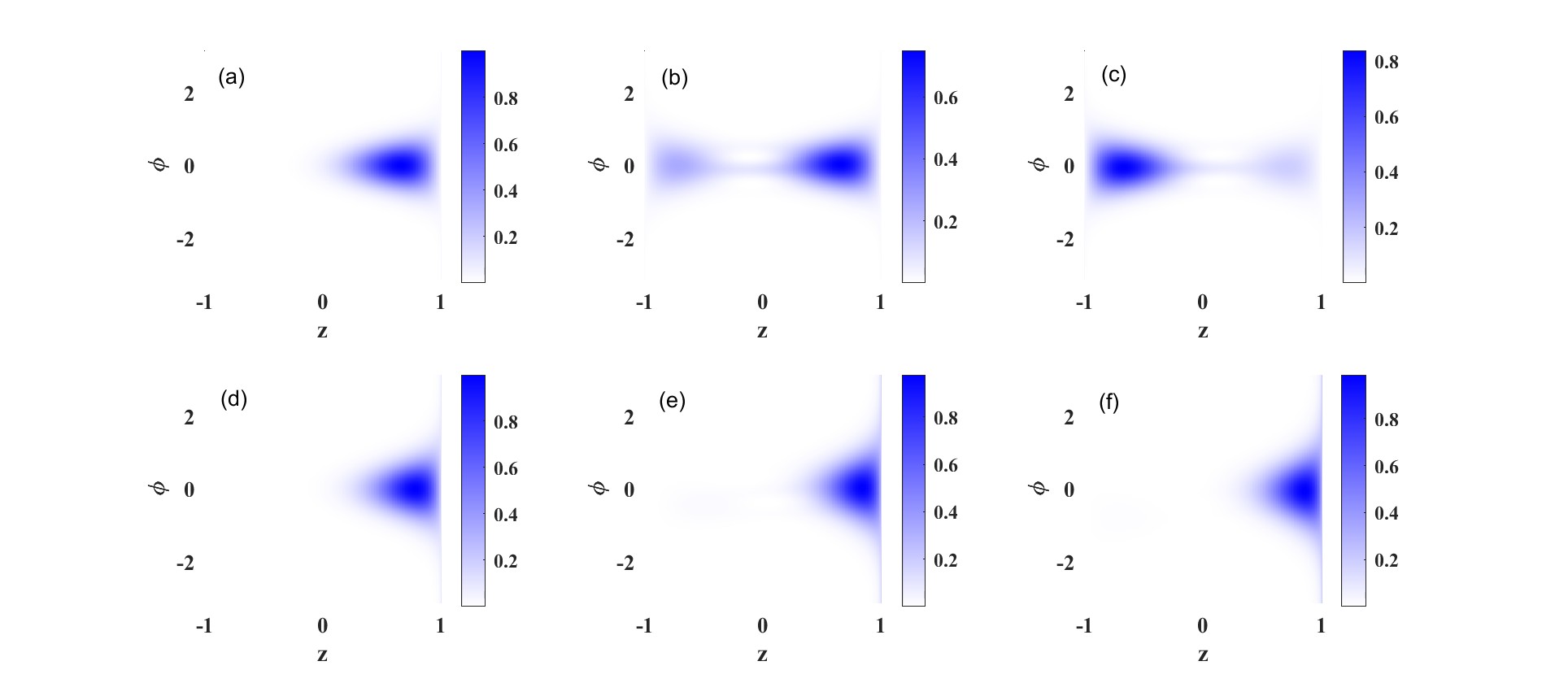}
\caption{Snapshots of beyond mean-field ($N=2$)
Husimi distributions for $S=20$ and different on-site coupling strengths and times:
(a)-(c): $U/J=-0.15$ and $t=0$ (a) $t=200$ (b) and $t=300$ (c);
(d)-(f): $U/J=-0.19$ and $t=0$ (d) $t=200$ (e) and $t=300$ (f).
The initial condition for $z$ is $z^{\rm SSB}-0.05$ with $z^{\rm SSB}
=0.71$ for panels (a)-(c) and  $z^{\rm SSB}
=0.83$ for panels (d)-(f).}\label{fig:husimi}
\end{figure}

The case of smaller particle number $S=20$ and $U/J=-0.15$
leads to $z^{\rm SSB}\approx 0.71$. The on-site interaction parameter
lies just between the classically predicted onset of SSB and the quantum
prediction. In the quantum case, it was shown that the SSB effect comes
along with the switch from a unimodal to a bimodal distribution of the
amplitudes in a
Fock space expansion of the ground state of the Hamiltonian \cite{MSPT11},
which for large $|U|$ becomes a so-called Schr\"odinger cat (NOON) state,
in our notation a superposition  proportional to $|S,0\rangle+|0,S\rangle$.
For smaller particle numbers, the onset of this effect, compared to the
mean-field prediction, is pushed to larger absolute values of $U$
(i.\ e., stronger attractive interaction), as shown
in Fig.\ 4 of \cite{Wim21}.
Thus for the parameters mentioned above, the phase space trajectories
in the beyond mean-field case show a much
different behavior than in the classical case.
To visualize this behavior, we have calculated the Husimi transform \cite{TWK09}
\begin{equation}
Q(z,\phi)=\langle\Omega|\Psi\rangle\langle\Psi|\Omega\rangle,
\end{equation}
with $|\Omega\rangle=|S,z,\phi\rangle$. This function is localized if the
time-evolved quantum state is localized around the stable fixed point, whereas
it is delocalized otherwise. Taking snapshots of its dynamics for $U/J=-0.15$,
displayed in panels (a) to (c) of Fig.\ \ref{fig:husimi}, a delocalization
of the dynamics can be observed, which is  very different from the
mean-field prediction, which is already deep in the SSB regime.
Using just two ACS, we can thus unravel this quantum effect
dynamically, without having to calculate the exact ground state.
Increasing the absolute value of the onsite interaction to
$|U|/J=0.19$, according to Eq.\ (\ref{eq:SSB}), the new stable fixed points are
moving towards larger values of $|z|$  and we are choosing an initial
condition with a small displacement away from the one with $\phi=0$ and a
positive value of $z$. The beyond mean-field result (with $N=2$) now also shows
a restriction of the dynamics to positive values of $z$, i.e., it displays
the phenomenon of SSB. This fact can be observed in panels (d) to (f) of
Fig.\ \ref{fig:husimi}. The fact that there is no motion from right to left
any more when $|U|$ is large, is due to the large energy barrier that has to be overcome in order to go from $z\approx 1$, i.e., approximately the state
$|S,0\rangle$ to $z\approx -1$, i.e., approximately the state $|0,S\rangle$ \cite{zhai_2021}.

\begin{figure}[h]
\centering
\includegraphics[width=0.49\textwidth]{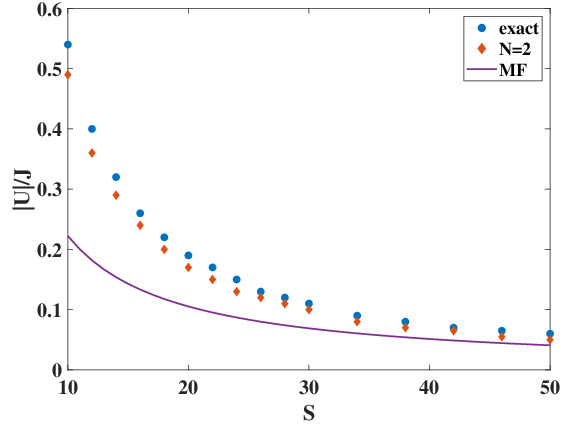}
\caption{Comparison of the onset of SSB as a function of $|U|/J$
for different particle number as predicted by: (i) ACS mean field (solid line),
(ii) multi-configuration ACS with $N=2$ (red diamonds) and
(iii) exact result (blue dots), inferred from bimodality of ground state.}
\label{fig:hyper}
\end{figure}

If one just wants to determine the transition from
localized motion to delocalized motion beyond the mean-field prediction,
it turned out above that only two ACS trajectories might be enough.
We will show in the remainder of this section that in order to almost
faithfully predict the occurence of the SSB transition in terms of $|U|$
for different particle numbers, a multiplicity of $N=2$ is indeed sufficient.
To show this, in Fig.\ \ref{fig:hyper}, we display the onset of SSB for
values of $S$, as predicted
by mean field (using a single ACS), yielding the hyperbolic dependence
$|U|/2J=1/(S-1)$ depicted by the solid line, to the result for this onset from
a multi-configurational calculation with $N=2$ (red diamonds).
To determine the location of the red diamonds, we have propagated the dynamics
for large enough times ($Jt=1000$) to be sure that the motion is either
confined to the right hand side of phase space (i.e., $z>0$) or not and have used
an interval nesting strategy to determine the onset of SSB. These two results are
then compared to the exact quantum ones (blue dots), calculated by monitoring the
expansion of the ground state (GS) in terms of Fock states. If the magnitude of
coefficients shows a bimodal structure, and $\langle {\rm GS}|\frac{S}{2},\frac{S}{2}\rangle\approx 0$, the SSB range is reached \cite{MSPT11}.
The multi-configuration ACS calculations with $N=2$ show a surpringly good
agreement with the exact quantum results, even for small particle numbers.

\section{Conclusions and Outlook}

We have reinvestigated some well-known physical phenomena in the
dynamics of the bosonic Josephson junction model using a powerful
multi-configuration technique to solve the TDSE. It was shown that,
by use of an expansion of the wave-function in multiple  ACSs, a
decisive improvement of the classical mean-field results
towards full quantum results can be achieved.
While in a Fock space calculation, the full basis has always to be
used, in the present approach, the size of the time-dependent basis
can be increased in order to achieve convergence and to reveal
quantum effects. The equations of motion
for the (time-dependent) variational parameters as well as for the
expansion coefficients are derived from the time-dependent
variational principle. This technical aspect of the
presented work is similar in spirit to the variational solution of
the Gross-Pitaevskii equation with long range interactions, based on
Gaussian wavepackets (Glauber coherent states) \cite{RMW10} as well
as to the multi-configurational time-dependent Hartree-Fock method
for bosons \cite{ASC08}, although in the latter case, the employed
basis functions are orthogonal. Furthermore, in contrast to the
Glauber coherent states, the ACS used herein conserve particle
number and are thus considered to be favorable in the
present case \cite{TWK09}. In addition, we stress that in contrast
to semiclassical methods that are based on Monte Carlo sampling of
initial conditions for mean-field trajectories and require
around 10$^4$ samples, here we can get satisfactory results with
only a handful of variationally determined ``trajectories''.
The semiclassical method employed by Tomsovic et al \cite{TPUUR18},
requires an order of magnitude less mean-field trajectories
(even in the 6 well case)
than the initial value semiclassical method refered to above,
but one has to find saddle points in a complexified phase space,
which is a formidable task.

The parameter space that we have covered is characterized by the strength
(and the sign) of the on-site interaction as well as by the total particle
number and the initial population imbalance. Firstly, by taking into
account one additional ACS, i.\ e., by employing a total of just two ACS,
the beating of the population imbalance (as well
as of the expectation of the sine of the phase operator) for small
positive values of $U$ can be
reproduced almost quantitatively exactly, if the initial imbalance is
rather small, i.\ e., if it is close to the classical equilibrium point at
the origin of phase space. The choice of the initial phase
variable of the second ACS was crucial to achieve this agreement.
For larger initial imbalance, the number of ACS needed to achieve reasonable
agreement with the exact quantum results has to be increased, with more and
more states needed, the higher the total particle number.

Secondly, our focus was on the more demanding parameter regime of MQST.
There, we could show that the use of more than ten ACS is necessary,
if the quantum reduction compared to the mean-field value of the repulsive interaction strength at which MQST sets in is to be uncovered. As had been noticed
before by Wimberger et al. \cite{Wim21}, a single ACS is not enough
to observe this effect. In the case of higher multiplicities $N>2$,
the choice of initial conditions for those
ACS that are initially unpopulated (i., e., the ones, whose
coefficients $A_k$ in Eq.\ (\ref{multi_su2}) are zero) was done by the random
sampling described in \cite{pra21}.

Lastly, for negative values of the on-site interaction
and for large particle numbers, we observed a beating oscillation
around the symmetry breaking equilibrium point, that still resembles
the mean-field trajectory, with the only quantum effect being the spiraling
in and out of the phase space trajectory. For small particle numbers,
compared to the mean-field prediction, the symmetry breaking only occurs
for larger attractive interaction in the quantum case, however \cite{Wim21}.
The fact that the symmetry breaking is lost for parameters that
would allow for symmetry breaking in mean-field theory is uncovered by
using just two ACS. The new prediction of the onset of symmetry
breaking in Fig.\ \ref{fig:hyper} is very close to the exact quantum result.

In future works, the fact that the addition of only a few
generalized coherent state
basis functions allows for the unraveling of quantum effects
can be put to good use. A possible extension of the present work would
be keeping the site number at two but allowing for more than
just a single atomic species \cite{Dufetal17}.
Furthermore, driven bosonic Josephson junctions show
dynamical tunneling \cite{GeHo14}
and the addition of a decay term in one of the sites allows
for a  characteristic modulation of the self-trapping \cite{GKN08}.
The description of these effects beyond mean field is a worthwhile
topic of future investigations.
Finally, if one also lets the site number $M$ grow, it might be the
only possibility
to use flexible time-dependent GCS basis functions if numerical results
showing quantum effects are asked for. This is due to the fact that the
number of Fock state basis functions grows like $\frac{(M+S-1)!}{S!(M-1)!}$
and Fock-state-based calculations thus become unfeasible.

\section{Methods}

\subsection{Linearized mean-field equations and their solution}
\label{app:plasma}

From the Jacobi  matrix in Eq.\ (\ref{eq:Jac1})
we read off the linearized equations of motion
\begin{eqnarray}
\dot{z}&=&2J\phi,
 \\
\dot{\phi}&=&-\left[2J+U(S-1)\right]z
\end{eqnarray}
for the population imbalance and the phase difference,
valid around the phase space origin.
Employing the initial conditions $z(0)=z_0, \phi(0)=0$,
their solution is given by
\begin{eqnarray}
 z(t)&=&\frac{z_0}{2}\left[e^{2J\sqrt{-(1+\Lambda)}t}+e^{-2J\sqrt{-(1+\Lambda)}t}\right],
 \\
 \phi(t)&=&\frac{z_0}{2}\sqrt{-(1+\Lambda)}\left[e^{2J\sqrt{-(1+\Lambda)}t}-e^{-2J\sqrt{-(1+\Lambda)}t}\right],
\end{eqnarray}
with the strength parameter $\Lambda$, defined in
Eq.\ (\ref{eq:Lam}) of the main text.

If $\Lambda>-1$, we have the oscillatory solutions
\begin{eqnarray}
 z(t)&=&z_0\cos(\Omega t)
 \\
 \phi(t)&=&-z_0\sqrt{1+\Lambda}\sin(\Omega t)
\end{eqnarray}
with the plasma frequency
\begin{equation}
 \Omega=2J\sqrt{1+\Lambda}.
 \end{equation}
We stress that for the specific choice of initial condition,
the oscillation amplitude of $\phi$ depends on the
strength parameter, while that of $z$ does not.

If $\Lambda<-1$, we obtain
\begin{eqnarray}
 z(t)&=&{z_0}\cosh\left[2J\sqrt{-(1+\Lambda)}t\right],
 \\
 \phi(t)&=&{z_0}\sqrt{-(1+\Lambda)}\sinh\left[2J\sqrt{-(1+\Lambda)}t\right],
\end{eqnarray}
which describes a solution like the ones displayed in panel
(b) of Fig.\ \ref{fig:mf}, but only
where the conditions $|z|,|\phi|\ll1$
are still fulfilled. Away from that regime the hyperbolic solution
is unphysical.

\subsection{Exact quantum calculation}\label{app:exact}

For the exact quantum results, we employ an expansion of the
wave-function in terms of Fock states
 \begin{equation}
 |\Psi(t)\rangle=\sum_{i=0}^{S}b_i(t)|F_i\rangle,
 \end{equation}
where the sum is taken over all the states $\{|F_i\rangle\}$ that
emerge if a total of $S$ particles is distributed over two sites.
Due to the fact that one can place from zero up to $S$ particles in,
e.g., the first site, it is obvious that there are $S+1$ different
possibilities.

In order to completely specify the
problem, the initial state has to be known, from which the $b$ coefficients
at $t=0$ can be extracted. In the present work, we consider an initial state
that is given in terms of a single ACS with parameters $\xi_1$ and $\xi_2$.
From the definition given in Eq.\ (\ref{eq:GCS}) taken for $M=2$,
by applying the binomial theorem, due to
$(\hat a_i^\dagger)^n|0\rangle=\sqrt{n!}|n\rangle$, we find
 \begin{eqnarray}
 |\Psi(0)\rangle&=&\frac{1}{\sqrt{S!}}(\xi_1\hat a_1^\dagger+
                                    \xi_2\hat a_2^\dagger)^S|0,0\rangle
                                    \nonumber\\
                &=&
                \sum_{j=0}^S
                \sqrt{\frac{S!}{{(S-j)!j!}}}
                \xi_1^{S-j}\xi_2^{j}|S-j,j\rangle,
 \end{eqnarray}
which is the Fock state expansion of the initial state, providing us with
the sought for coefficients at $t=0$.

The first option to evolve the wave-function over time would be to solve the coupled system of linear differential equations for the $b$ coefficients
\begin{equation}
 {\rm i}\dot b_j(t)=\sum_{i=0}^S\langle F_j|\hat H|F_i\rangle b_i(t),
\end{equation}
that follows from the TDSE, e.\ g., by using a Runge-Kutta method or by
matrix exponentiation (which in the present case of time-independent
Hamiltonian turns out to be advantageous, because the matrix exponential
has to be calculated only once, before the propagation loop is started).
An alternative, second option, which is
also numerically exact, would require diagonalising
the BH Hamiltonian \cite{TLF05}, e.g., in the Fock basis,
see also Section 2.3.1 in \cite{Gross3}.
The time evolution is then finally given by ($\hbar=1$)
  \begin{equation}
 |\Psi(t)\rangle=\sum_{i=0}^{S}c_i\exp\{-{\rm i}E_it\}|\Phi_i\rangle,
 \end{equation}
where  $\{E_i\}$ are the eigenenergies and the $\{|\Phi_i\rangle\}$ are
the eigenstates. The time-independent $c$-coefficients follow from
the expansion of the initial wave-function in the eigenstates.

In both cases, the matrix elements of the Hamiltonian have to be set up.
This does not pose a major challenge in case of small site numbers but
in the general case it requires some clever way of creating and labeling
of the Fock states, as described in a pedagogical way in \cite{ZhDo10}.

\section*{Conflict of Interest Statement}

The authors declare that the research was conducted in the absence of any commercial or financial relationships that could be construed as a potential conflict of interest.

\section*{Author Contributions}

Yulong Qiao performed the numerical studies presented in this work.
Both authors contributed to the conception of the research and to the writing of the manuscript.

\section*{Acknowledgments}
The authors would like to thank Prof.\ A.\ R.\ Kolovsky for computer code to perform the exact Fock space calculations (using the first option mentioned in Section 5.2) whose results were displayed here.

\bibliography{lib_bohu}


\end{document}